\documentstyle[]{article}

\title
{
{\small In
C. Freksa, ed.,
{\em 
Foundations of Computer Science: Potential - Theory - Cognition}\\
\vspace{-.3cm}
Lecture Notes in Computer Science, pp. 201-208, Springer, 1997. } \\
\vspace{.5cm}
A Computer Scientist's View of \\
Life, the Universe, and Everything \\
}

\author{J\"{u}rgen Schmidhuber \\
IDSIA, Corso Elvezia 36, CH-6900-Lugano, Switzerland \\
{\tt juergen@idsia.ch - http://www.idsia.ch/\~{ }juergen} 
}

\date{}

\begin{document}

\maketitle

\begin{abstract}

Is the universe computable?  If so, it may be much cheaper in terms of
information requirements to compute all computable universes instead
of just ours.  I apply basic concepts of Kolmogorov complexity theory
to the set of possible  universes, and chat about perceived and true
randomness, life, generalization, and learning in a given universe.

\end{abstract}

\section*{{\sc Preliminaries}}

\hspace{0.4cm}
{\bf Assumptions.}
A long time ago, the Great Programmer wrote a program that
runs all possible universes on His Big Computer.
``Possible'' means ``computable'':
(1) Each universe evolves on a discrete time scale. 
(2) Any universe's state at a given time is
describable by a finite number of bits.
One of the many universes is ours, despite some who evolved 
in it and claim it is incomputable.

{\bf Computable universes.}
Let $TM$ denote an arbitrary universal Turing machine  
with unidirectional output tape. $TM$'s
input and output symbols are ``0'', ``1'', and ``,'' (comma).  
$TM$'s possible input programs can be ordered alphabetically:
``'' (empty program),
``0'',
``1'',
``,'',
``00'',
``01'',
``0,'',
``10'',
``11'',
``1,'',
``,0'',
``,1'',
``,,'',
``000'', etc.
Let $A_k$ denote $TM$'s $k$-th program in this list.
Its output will be a finite or infinite string over the alphabet
$\{$ ``0'',``1'',``,''$\}$. This 
sequence of bitstrings separated by commas
will be interpreted as
the evolution $E_k$ 
of universe $U_k$.
If $E_k$ includes at least one comma,
then let $U_k^l$ 
denote the $l$-th (possibly empty) bitstring 
before the $l$-th comma.
$U_k^l$ represents $U_k$'s state at the $l$-th time step of $E_k$
($k, l \in \{1, 2, \ldots, \}$).
$E_k$ is represented
by the sequence $U_k^1, U_k^2,  \ldots$ 
where $U_k^1$ corresponds to $U_k$'s big bang.
Different algorithms may compute the same universe.
Some universes are finite (those whose programs
cease producing outputs at some point), others are not.
I don't know about ours.

{\bf TM not important.}
The choice of the Turing machine is not important.
This is due to the compiler theorem:
for each universal Turing machine $C$ there exists a constant prefix $\mu_C$
$\in \{$ ``0'',``1'',``,''$\}^*$
such that for all possible programs $p$, $C$'s output in response to  
program $\mu_C p$ is identical to
$TM$'s output in response to $p$.
The prefix $\mu_C$ is the compiler that compiles programs for $TM$ into
equivalent programs for $C$.

{\bf Computing all universes.}
One way of sequentially computing all computable universes is 
dove-tailing. $A_1$ is run for one instruction every second step, $A_2$ is
run for one instruction every second of the remaining steps, and so on.
Similar methods exist for computing many universes in parallel.
Each time step of each  universe that is computable by at least one 
finite algorithm will eventually be computed.

{\bf Time.}
The Great Programmer does not worry about computation time. 
Nobody presses Him. 
Creatures which evolve in any of the universes don't have to worry 
either. They run on local time and have no idea of how many instructions
it takes the Big Computer to compute one of their time steps, or how many
instructions it spends on all the other creatures in parallel universes.  

\section*{{\sc Regular and Irregular Universes }}

\hspace{0.4cm}
{\bf Finite histories.}
Let $\mid x \mid$ denote the number of symbols in string $x$.
Let the partial history $S_k^{i,j}$ denote the substring between 
the $i$-th and the $j$-th symbol of $E_k$, $j > i$.
$S_k^{i,j}$ is regular (or compressible, or non-random)
if the shortest program that computes $S_k^{i,j}$
(and nothing else)
{\em and halts} consists of less than $ \mid S_k^{i,j} \mid $ symbols.
Otherwise $S_k^{i,j}$ is irregular (incompressible, random).

{\bf Infinite histories.}
Similarly,
if some universe's evolution is infinite, 
then it is compressible if it can be computed by a finite algorithm.

{\bf Most universes are irregular.}
The evolutions of almost all universes are incompressible.
There are $3^n$ strings of size $n$, but less than
$(1/3)^c * 3^{n} << 3^n$ algorithms
consisting of less than $n-c$ symbols 
($c$ is a positive integer).
And for the infinite case, we observe:
the number of infinite symbol strings is incountable.
Only a negligible fraction (namely countably many of them)
can be computed by finite programs.

{\bf The few regular universes.}
There are a few compressible universes which can be computed by very
short algorithms, though. For instance, suppose that some $U_k$ evolves according 
to physical laws that tell us how to compute next states from previous states.
All we need to compute $U_k$'s evolution is  $U_k^1$ and the algorithm that computes  
$U_k^{i+1}$ from $U_k^i$ ($i \in \{1, 2, \ldots, \}$). 

{\bf Noise?}
Apparently, we live in one of the few  highly regular universes.
Each electron appears to behave the same way. 
Dropped breads of butter regularly hit the floor, not the ceiling. 
There appear to be deviations from regularity, however, 
embodied by what we call noise.
Although certain macroscopic properties (such as pressure in a 
gas container) are predictable by physicists,
microscopic properties (such as precise particle positions) seem
subject to noisy fluctuations. Noise represents additional information
absent in the original physical laws.  Uniform noise is 
incompressible --- there is no short algorithm that computes it and
nothing else.

{\bf Noise does not necessarily prevent compressibility.}
Laws currently used by physicists to model our own universe model
noise.  Based on Schr\"{o}dinger's equation,
they are  only conditional probability 
distributions on possible next states, given previous states. 
The evolution of Schr\"{o}dinger's wave function (WF) itself can be computed by
a very compact algorithm (given the quantizability assumptions in the first
paragraph of this paper) --- WF is just a short formula.
Whenever WF collapses in a particular way, however, the
resulting actual state represents additional information (noise) not conveyed
by the algorithm describing the initial state (big bang) and WF. 
Still, since the noise 
obviously is non-uniform (due to the nature of the physical laws and
WF), our universe's evolution so far is greatly compressible. How? 
Well, there is a comparatively short algorithm that 
simply codes probable next states by few bits, 
and unlikely next states by many bits, 
as suggested by standard information 
theory \cite{Shannon:48}.

{\bf More regularity than we think?}
The longer the shortest program computing a given universe, 
the more random it is.
To certain observers, certain universes appear partly 
random although they aren't. 
There may be at least two reasons for this:

{\bf 1. Shortest algorithm cannot be found.}
It can be shown that
there is no algorithm that can generate the shortest
program for computing arbitrary given data on a given computer
\cite{Kolmogorov:65,Solomonoff:64,Chaitin:87}.
In particular, our physicists cannot expect to find the 
most compact description of our universe.

{\bf 2. Additional problems of the Heisenberg type.}
Heisenberg tells us that we cannot even observe the precise, current state of a 
single electron, let alone our universe. In our particular
universe, our actions seem to influence our measurements in a fundamentally
unpredictable way.
This does not mean that there is no predictable 
underlying computational process (whose precise results we cannot access).
In fact, rules that hold for observers who are 
part of a given universe and evolved according to its 
laws need not hold outside of it.  There is no reason to believe that
the Great Programmer cannot dump a universe and examine its precise state at any 
given time, just because the creatures that evolved in it cannot because their 
measurements modify their world. 

{\bf How much true randomness?}
Is there ``true'' randomness in our universe, 
in addition to the simple physical laws? 
True randomness essentially means 
that there is no short algorithm computing ``the precise collapse of the
wave function'', and what is perceived as noise by today's physicists.
In fact, if our universe was infinite, and there was true randomness,
then it could not be computed by a finite algorithm that computes nothing
else.  Our fundamental inability to perceive 
our universe's state does {\em not} imply its true randomness, though. 
For instance, there may be a very short algorithm computing the positions 
of electrons lightyears apart in a way that seems like noise to 
us but actually is highly regular.  

\section*{{\sc  All Universes are Cheaper Than Just One}}

In general, computing all evolutions of all universes is
much cheaper in terms of information requirements than computing just 
one particular, arbitrarily chosen evolution. Why?
Because the Great Programmer's algorithm that systematically 
enumerates and runs all universes
(with all imaginable types of physical laws, wave functions, noise etc.)
is {\em very} short (although it takes time). 
On the other hand,
computing just one particular universe's  evolution (with, say, one particular
instance of noise), without computing the others,
tends to be very expensive, 
because almost all individual universes are 
incompressible, as has been shown above. 
More is less!

{\bf Many worlds.}
Suppose there is true (incompressible) noise
in state transitions of our particular world evolution.
The noise conveys additional information 
besides the one for initial state and physical laws.  
But from the Great Programmer's point of view, almost no extra
information (nor, equivalently, a random generator) is required.
Instead of computing just one of the many possible evolutions of 
a probabilistic universe with fixed laws but random noise of a 
certain (e.g., Gaussian)  type, the Great Programmer's 
simple program computes them all. 
An automatic  by-product of the Great Programmer's set-up is
the well-known ``many worlds hypothesis'', \copyright  Everett III.
According to it, whenever our 
universe's quantum mechanics allows for alternative next paths, 
all are taken and the world splits into separate universes. 
From the Great Programmer's view, however,
there are no real splits --- there are just a bunch of  
different algorithms which yield identical results for some time, until 
they start computing different outputs corresponding to different noise 
in different universes.

From an esthetical point of view that favors simple explanations
of everything, a set-up in which all possible
universes are computed instead of just ours
is more attractive. It is simpler.

\section*{{\sc Are we Run by a Short Algorithm? }}

Since our universes' history so far is regular,
it by itself {\em could} have been computed by a 
relatively short algorithm.
Essentially, this algorithm embodies the physical 
laws plus the information about the historical noise. 
But there are many algorithms whose output sequences start with our 
universe's history so far. Most of them are very long.
How likely is it now that our universe is indeed run by a short algorithm?
To attempt an answer, we need a prior probability on the possible 
algorithms.  The obvious candidate  is the ``universal prior''.

{\bf Universal prior.}
Define $P_U(s)$, the {\em a priori probability} of a finite symbol 
string $s$ (such as the one representing our universe's history so far),
as the probability of guessing a halting program
that computes $s$ on a universal Turing machine
$U$. Here, the way of guessing is defined
by the following procedure:
initially, the input tape consists of a single square.
Whenever the scanning head of the program tape shifts
to the right, do: (1) Append a new square.
(2) With probability $\frac{1}{3}$ fill it with a ``0'';
with probability $\frac{1}{3}$ fill it with a ``1'';
with probability $\frac{1}{3}$ fill it with a ``,''.
Programs are ``self-delimiting'' 
\cite{Levin:74,Chaitin:87}
--- once $U$ halts due
to computations based on
the randomly chosen symbols (the program) on its input tape,
there won't be any additional program symbols.
We obtain
\[
P_U(s) = \sum_{p: U~computes~s~from~p~and~halts} (\frac{1}{3})^{\mid p \mid}.
\]
Clearly,
the sum of all probabilities of all halting programs cannot exceed 1
(no halting program can be the prefix of another one).
But certain programs may lead to
non-halting computations.

Under different universal priors (based on different universal
machines), probabilities of a given string
differ by no more than a constant factor independent of the
string size, due to the compiler 
theorem (the constant factor corresponds to the probability of
guessing a compiler).
This justifies the name ``{\em universal}  prior,''
also known as Solomonoff-Levin distribution.

{\bf Dominance of shortest programs.}
It can be shown (the proof is non-trivial) that
the probability of
guessing any of the programs computing some string
and the probability of
guessing one of its shortest programs
are essentially equal (they differ by no more than a constant factor
depending on the particular Turing machine).
The probability of a string is dominated by the probabilities of its
shortest programs. This is known as the ``coding theorem''  \cite{Levin:74}. 
Similar coding theorems exist for the case of non-halting programs
which cease requesting additional input symbols at a certain point.

Now back to our question: are we run by a relatively compact algorithm?
So far our universe {\em could} have been run by one --- 
its history {\em could} have been much noisier and thus much less
compressible. Hence universal prior and coding theorems suggest that the
algorithm is indeed short. If it is, then there will be less than
maximal randomness in our future, and more than vanishing predictability.
We may hope that our universe will remain regular, as opposed to
drifting off into irregularity.

\section*{{\sc Life in a Given Universe}}

\hspace{0.4cm}
{\bf Recognizing life.}
What is life? The answer depends on the observer.
For instance, certain substrings of $E_k$ may be interpretable
as the life of a living thing $L_k$ in $U_k$. 
Different observers will have different views, though. What's life to
one observer will be noise to another. In particular, if the observer
is not like the Great Programmer but
also inhabits $U_k$, then its own life may 
be representable by a similar substring.
Assuming that recognition implies relating observations to previous knowledge,
both  $L_k$'s and the observer's life will have to
share mutual algorithmic information \cite{Chaitin:87}:
there will be a comparatively
short algorithm computing $L_k$'s from the observer's life, and vice versa.

Of course, creatures living in a given universe don't
have to have any idea of the symbol strings by which they
are represented. 

{\bf Possible limitations of the Great Programmer.}
He does not need not be very smart. For instance, in some of His
universes phenomena will appear that humans would call life.
The Great Programmer won't have to be able to recognize them. 

{\bf The Great Programmer reappears.}
Several of the Great Programmer's universes 
will feature another Great Programmer who programs another Big Computer 
to run all possible universes. Obviously, there are infinite chains of
Great Programmers. If our own universe allowed for enough storage,
enough time, and fault-free computing, then you could be one of them.

\section*{{\sc Generalization and Learning}}

\hspace{0.4cm}
{\bf In general, generalization is impossible.}
Given the history of a particular universe up to a given
time, there are infinitely many possible continuations.
Most of these continuations have nothing to do with the previous history.
To see this, suppose we have observed 
a partial history $S_k^{i,j}$ (the substring between
the $i$-th and the $j$-th symbol of $E_k$).
Now we want to generalize from previous
experience to predict 
$S_k^{j+1,l}$, $l > j$.
To do this, we need an algorithm that computes $S_k^{j+1,l}$ from $S_k^{i,j}$
($S_k^{i,j}$ may be stored on a separate, additional
input tape for an appropriate
universal Turing machine).
There are $3^{l - j}$ possible futures.
But for $c < l - j$, there are less than $(1/3)^c * 3^{l - j}$ algorithms
with less than $l - j - c$ bits computing such a future,
given $S_k^{i,j}$.
Hence in most cases
the shortest algorithm computing the future, given the past, 
won't be much shorter than the
shortest algorithm computing the future from nothing.
Both will have about the size of the entire future. 
In other words, the mutual algorithmic information between history 
and future will be zero.
As a consequence, in most universes (those that can be computed by 
long algorithms only), successful generalization from previous experience is 
not possible. Neither is inductive transfer.
This simple insight is related to
results in \cite{Wolpert:96}.

{\bf Learning.}
Given the above, since learning means to use previous experience
to improve future performance, learning is possible only in 
the few regular universes 
(no learning without compressibility).
On the other hand, regularity by itself is not sufficient
to allow for learning. For instance, there is a highly compressible and
regular universe represented
by ``,,,,,,,...''. It is too simple to allow for processes we would
be willing to call learning.

In what follows, I will assume that a regular universe 
is complex enough to allow for 
identifying certain permanent data structures of a general learner 
to be described below.  For convenience,
I will abstract from bitstring models, and instead talk about
environments, rewards, stacks etc. Of course, all these abstract
concepts are representable as bitstrings.

{\bf Scenario.}
In general, the learner's life is limited. 
To it, time will be important (not to the Great Programmer though).
Suppose its life in environment $\cal E$
lasts from time 0 to unknown time $T$. 
In between it repeats the following cycle
over and over again ($\cal A$  denotes a set of possible actions):
select and execute $a \in \cal A$ with probability $P( a \mid \cal E )$,
where the modifiable policy $P$ is a variable,
conditional probability distribution on the possible actions,
given current $\cal E$.
Action $a$ will consume time and may change $\cal E$ and $P$.
Actions that modify $P$ are called primitive learning algorithms (PLAs). 
$P$ influences the way $P$ is modified (``self-modification'').
{\em Policy modification processes} (PMPs)
are action subsequences that include PLAs.
The $i$-th PMP in system life is denoted {\em PMP}$_i$,
starts at time $s_i > 0$,  ends at $e_i < T$, $e_i > s_i$,
and computes a sequence of $P$-modifications denoted $M_i$.
Both $s_i$ and $e_i$ are computed dynamically
by special instructions in $\cal A$
executed according to $P$ itself: $P$ says when to start and end PMPs. 

Occasionally $\cal E$ provides
real-valued reward.  The cumulative reward obtained in between
time 0 and time $t > 0$ is denoted by $R(t)$ (where $R(0) = 0$).
At each PMP-start $s_i$ the learner's goal is to
use experience to generate $P$-modifications to
accelerate future reward intake. 
Assuming that reward acceleration is possible at all,
given $E$ and $\cal A$, how can the learner achieve it? 
I will describe a rather general way of doing so.

{\bf The success-story criterion.}
Each PMP-start time $s_i$ will trigger an evaluation 
of the system's performance so far.  
Since $s_i$ is computed according to $P$,
$P$ incorporates information
about when to evaluate itself.
Evaluations may cause
policy modifications to be undone (by restoring
the previous policy --- in practical implementations,
this requires to store previous values of modified
policy components on a stack).
At a given PMP-start $t$ in the learner's life,
let $V(t)$ denot the set of those previous $s_i$ whose
corresponding $M_i$ have not
been undone yet.  If $V(t)$ is not empty,
then let $v_i ~~ (i \in \{1, 2, \ldots, \mid V(t) \mid \}$ 
denote the $i$-th such time, ordered according to size.
The success-story criterion
SSC is satisfied if either $V(t)$ is empty (trivial case) or if
\[
\frac{R(t)}{t}
<
\frac{R(t) - R(v_1)}{t - v_1}
<
\frac{R(t) - R(v_2)}{t - v_2}
<
\ldots
<
\frac{R(t) - R(v_{ \mid V(t) \mid }) }{t - v_{\mid V(t) \mid}}.
\]
SSC essentially says that
each surviving $P$-modification corresponds
to a long term reward acceleration.
Once SSC is satisfied, the learner 
continues to act and learn
until the next PMP-start.
Since there may be arbitrary reward delays in response
to certain action subsequences, it is important that
$\cal A$  indeed includes actions for
delaying performance evaluations --- the learner will have
to learn when to trigger evaluations.
Since the success of a policy modification recursively depends on the
success of later modifications for which it is setting the stage,
the framework provides a basis for ``learning how to learn''.
Unlike with previous learning paradigms, the entire life is considered
for performance evaluations.  Experiments in 
\cite{Schmidhuber:97bias,Schmidhuber:97ssa}
show the paradigm's practical feasibility.  
For instance, in \cite{Schmidhuber:97bias}
$\cal A$ includes an extension of Levin search \cite{Levin:84}
for generating the PMPs.

\section*{{\sc Philosophy}}

\hspace{0.4cm}
{\bf Life after death.}
Members of certain religious sects expect resurrection of the dead in a paradise
where lions and lambs cuddle each other. There is a possible continuation
of our world where they will be right. In other possible continuations, however,
lambs will attack lions.

According to the computability-oriented view adopted in this paper,
life after death is a technological problem, not a religious one.
All that is necessary for some human's resurrection is to record his
defining parameters (such as brain connectivity and synapse properties etc.),
and then dump them into a large computing device computing an appropriate
virtual paradise. Similar things have been suggested by
various science fiction authors.
At the moment of this writing, neither appropriate
recording devices nor computers of sufficient size exist. There is
no fundamental reason, however, to believe that they won't exist in 
the future.  

{\bf Body and soul.}
More than 2000 years of European philosophy dealt 
with the distinction between body and soul.
The Great Programmer does not care. 
The processes that correspond to our brain firing patterns and the
sound waves they provoke during discussions about body and soul 
correspond to computable substrings of our universe's evolution. 
Bitstrings representing such talk may evolve in many universes. 
For instance, sound wave patterns representing notions 
such as body and soul and ``consciousness'' may be useful in 
everyday language of certain inhabitants of those universes.
From the view of the Great Programmer, though,  such 
bitstring subpatterns may be entirely irrelevant. 
There is no need for Him to load them with ``meaning''.

{\bf Talking about the incomputable.}
Although we live in a computable universe, we occasionally chat
about incomputable things, such as the halting probability  of a 
universal Turing machine (which is closely related to G\"{o}del's 
incompleteness theorem). And we sometimes discuss inconsistent
worlds in which, say, time travel is possible. Talk about such
worlds, however, does not violate the consistency of the processes
underlying it.

{\bf Conclusion.}
By stepping back and adopting the Great Programmer's point of view,
classic problems of philosophy go away.


\section*{{\sc Acknowledgments}}
Thanks to Christof Schmidhuber for interesting discussions on
wave functions, string theory, and possible universes.


\bibliographystyle{plain}

\end{document}